# Design of Linear Residual Generators for Combined Fault Detection and Estimation in Nonlinear Systems

Sunjeev Venkateswaran, Costas Kravaris*

Artie McFerrin Department of Chemical Engineering
Texas A&M University
College Station, TX 77843-3122, USA

*Abstract*:

A systematic method for the design of linear residual generators for combined fault detection and estimation in nonlinear systems is developed. The proposed residual generator is a linear functional observer built for an extended system that incorporates the fault dynamics from a linear exo-system, and in addition possesses disturbance-decoupling properties. Necessary and sufficient conditions for the existence of such residual generators for nonlinear systems are derived. As long as these conditions are satisfied, we obtain explicit design formulas for the residual generator. The results are illustrated through a chemical reactor case study, which demonstrates the effectiveness of the proposed methodology.

*Keywords*: Nonlinear system fault diagnosis, residual generators for nonlinear systems, functional observers for nonlinear systems

______________________________________________________________________
*author to whom correspondence should be addressed. Email: kravaris@mail.che.tamu.edu

# 1. Introduction

Fault detection and isolation is an area of critical significance for the monitoring and proper operation of processes, especially in the face of potential safety issues. But in addition to detection and isolation, it is also very important to be able to estimate the size of the fault, to determine if it is 100% or partial failure, and based on that, decide on the most appropriate course of action. In fact, combined detection, isolation and estimation can form the basis of fault tolerant control algorithms, to perform potential switchings to alternative control inputs, to preserve the operation of the process (Mhaskar et al., 2012; Du at al., 2023; Ranjan & Kravaris, 2024).

Fault diagnosis algorithms that involve combined detection + isolation + estimation, are often based on physical models instead of data-fitted empirical models, because of the need for a broad-range predictive capability in the face of potential large-size faults. The theory of model-based fault diagnosis for linear systems is quite mature (Ding, 2008; Frank, 1987), but there has also been significant progress for non-linear systems (De Persis & Isidori, 2001; Venkateswaran et al., 2022; Venkateswaran & Kravaris, 2023).

In model-based fault diagnosis, one typically builds a special type of functional observer, called residual generator, which gives a zero signal in the absence of a fault and a non-zero in the presence of a fault, and thus performs the detection of the fault. To perform isolation, one can use a bank of residual generators, one for each fault, that are decoupled from each other. In recent work of the authors (Venkateswaran et al., 2022), the possibility of building linear residual generators for detection and isolation in nonlinear systems has been investigated, including the derivation of necessary and sufficient feasibility conditions, as well as concrete design guidelines for the linear residual generators. The present work is a continuation of our efforts, where the residual generator must also provide an estimate of the fault size: the output of the residual generator must represent an estimate of the fault.

The methodology followed in the present paper will be in the same vein as in Venkateswaran et al. (2022) in the sense that the residual generator will still be a disturbance-decoupled functional observer, that directly estimates the fault itself instead of the baseline of zero. However, the resulting feasibility conditions will be somewhat more restrictive, and also there will be a requirement of a priori knowledge of the nature of the fault (whether it is a step, a ramp, etc.).

Section 2 will formulate the problem of building a linear residual generator that performs detection and estimation of a fault at the same time, and the design equations for the pertinent functional observer will be defined, including disturbance-decoupling specifications. Section 3 will derive necessary and sufficient conditions for the existence of a linear residual generator as defined in Section 2, including a concrete formula for the residual generator. The results will be compared to our previous results (Venkateswaran et al., 2022), where only detection was sought. Finally, in Section 4, the proposed method will be illustrated in a chemical reactor case study, where a safety-critical exothermic chemical reaction takes place and with potential faults in both the cooling system and the analytical sensor. Simulation results demonstrate the effectiveness of the proposed approach in detecting, isolating and estimating the faults.

## 2. Disturbance-decoupled detection and estimation using a linear functional observer as a residual generator

Consider a nonlinear process described by:

$$\dot{x} = F(x) + G(x)f + \sum_{i=1}^{m} E_i(x)\, w_i$$

$$y = H(x) + J(x)f + \sum_{i=1}^{m} K_i(x)\, w_i \tag{2.1}$$

where $x \in \mathbb{R}^n$ denotes the vector of states, $y \in \mathbb{R}^p$ denotes the vector of measured outputs. $f \in \mathbb{R}$ and $w_i \in \mathbb{R}, i = 1,2,\ldots,m$ are the fault and the disturbances/uncertainties respectively (system inputs) and $F(x), G(x), E_i(x), H(x), J(x), K_i(x)$ are smooth functions. Under normal operation of the process, the input $f$ (fault) is identically equal to zero, however in an abnormal situation (equipment failure), $f$ becomes nonzero, and this is what needs to be detected and estimated on the basis of the measurements. The inputs $w_i$ describe normal variability of process conditions (disturbances) and/or model uncertainty. It is in the presence of this variability that the fault must be detected and estimated, and the outcome must be unaffected by the presence of $w_i$ (disturbance-decoupled detection and estimation).

A fault is an unexcepted major deviation in process variables from the usual conditions. Faults may be categorized into different types based on their location. These include: (i) sensor faults (ii) component faults and (iii) actuator faults. Sensor faults may degrade performance of decision-making systems, including feedback control system, safety control system, quality control system, state estimation system, optimization system. The most common sensor faults include a) bias b) drift c) performance degradation d) sensor freezing e) calibration error (Ding, 2008). A fault in an actuator may cause loss of control in automated control systems. Actuator faults include, for example, stuck-up of control valves and faults in pumps, etc. Several common faults in servomotors include Lock-in-Place (LIP), Float, Hard-over Failure (HOF) and Loss of Effectiveness (LOE). Component faults occur in the equipment of plant. Examples for this could include leaks/blockages in pipeline and tanks. These faults change the physical characteristics of the component and as a result can lead to significant change in the dynamics of the process.

In this work, we consider faults generated from a linear exo-system of the following form:

$$\dot{x}_o = Rx_o \qquad (2.2)$$

$$f = Qx_o$$

where $x_o \in \mathbb{R}^{n_o}$, R and Q are $n_o \times n_o$ and $1 \times n_o$ matrices respectively. Examples of faults that can be represented by an exo-system (2.2) include step $(R = 0, Q = 1)$, ramp $(R = \begin{bmatrix} 0 & 1 \\ 0 & 0 \end{bmatrix}, Q = [1\ 0]$ and sine wave $R = \begin{bmatrix} 0 & \omega \\ -\omega & 0 \end{bmatrix}, Q = [1\ 0])$. This formulation is in the same vein as the one taken in the internal model principle in regulation problems to eliminate tracking errors in the presence of disturbances (Francis & Wonham, 1975; Francis & Wonham, 1976; Isidori et al., 2003). Equation (2.2) can be thought of as a disturbance generator (Isidori et al., 2003).

The overall system of process (2.1) and exo-system (2.2) in cascade is

$$\dot{x} = F(x) + G(x)Qx_o + \sum_{i=1}^{m} E_i(x) w_i$$

$$\dot{x}_o = Rx_o \qquad (2.3)$$

$$y = H(x) + J(x)Qx_o + \sum_{i=1}^{m} K_i(x) w_i$$

$$f = Qx_o$$

In this work, we will study the problem of disturbance-decoupled fault detection and estimation using a **linear** functional observer of order $s$ of the form

$$\dot{z} = Az + By \qquad (2.4)$$

$$\hat{f} = Cz + Dy$$

with state $z \in \mathbb{R}^s$, output $\hat{f} \in \mathbb{R}$ (the residual estimate), and parameters A, B, C, D being $s \times s$, $s \times p$, $1 \times s$ and $1 \times p$ matrices respectively with $(C, A)$ observable pair, that will be designed for system (2.3) so that the response of the estimate $\hat{f}$ of the series connection of (2.3) and (2.4) (see Figure 1)

$$\frac{d}{dt}\begin{bmatrix} x \\ x_o \\ z \end{bmatrix} = \begin{bmatrix} F(x) + G(x)Qx_o \\ Rx_o \\ Az + BH(x) + BJ(x)Qx_o \end{bmatrix} + \sum_{i=1}^{m} \begin{bmatrix} E_i(x) \\ 0 \\ BK_i(x) \end{bmatrix} w_i \qquad (2.5)$$

$$\hat{f} = [Cz + DH(x) + DJ(x)Qx_o] + \sum_{i=1}^{m}[DK_i(x)]w_i$$

has the following properties:

i. $\hat{f}$ asymptotically approaches $f = Qx_o$

ii. $\hat{f}$ is unaffected by the disturbances $w_i, i = 1, 2, \ldots, m$.

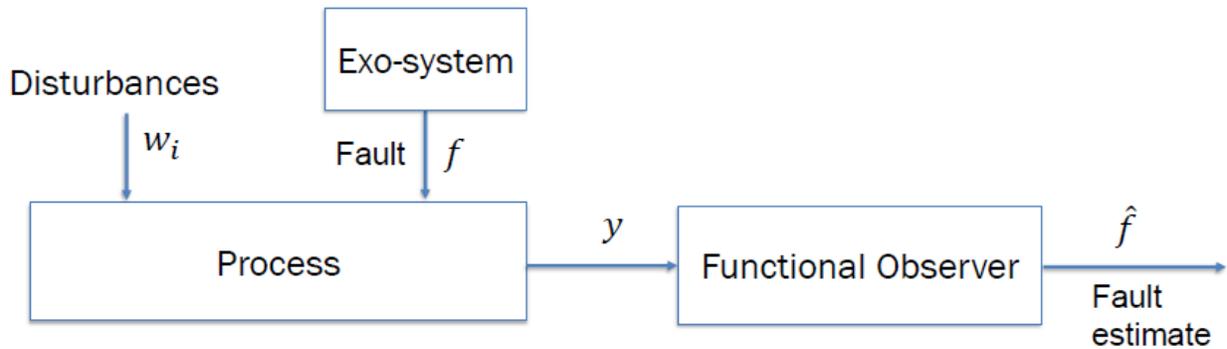

Figure 1. Fault Detection and Estimation Scheme

The choice of **linear** functional observers for nonlinear systems, whenever possible, has the advantage of practicality of linear filters in design and implementation. We will derive necessary and sufficient conditions for the existence of a linear residual generator based on a disturbance-decoupled linear functional observer. As long as these conditions are satisfied, we will derive simple design formulas for the residual generator, with eigenvalue assignment possibility.

*Functional observer design conditions for disturbance-decoupled fault estimation*

In this subsection we derive specific design conditions that the functional observer must satisfy to meet the requirements (i)-(ii). The first and foremost requirement of the functional observer is that in the absence of disturbances or initialization error, it must exactly reconstruct the fault: the fault estimation error must asymptotically converge to zero in the presence of initialization errors. Consequently (Luenberger, 1966; Luenberger, 1971; Ding, 2008; Kravaris, 2016; Kravaris & Venkateswaran, 2021) there must exist a differentiable map $T(x, x_o)$ from $\mathbb{R}^{n+n_o}$ to $\mathbb{R}^s$ such that:

$$\frac{\partial T(x, x_o)}{\partial x}(F(x) + G(x)Qx_o) + \frac{\partial T(x, x_o)}{\partial x_o}Rx_o = AT(x, x_o) + B(H(x) + J(x)Qx_o) \quad (2.6)$$

$$CT(x, x_o) + D\big((H(x) + J(x)Qx_o)\big) = Qx_o \quad (2.7)$$

Conditions (2.6) and (2.7) state that $z = T(x, x_o)$ is an invariant manifold of the zero-input dynamics of system (2.5), on which the fault estimate $\hat{f}$ is identically equal to $Qx_o = f$.

If conditions (2.6) and (2.7) are satisfied, the observer error dynamics (expressed in terms of the off-the-manifold coordinate $e = z - T(x, x_o)$ ) and the fault estimation error $\hat{f} - f$ are given by:

$$\frac{d(z - T(x, x_o))}{dt} = A\big(z - T(x, x_o)\big) + \sum_{i=1}^{m}\left(BK_i(x) - \frac{\partial T(x, x_o)}{\partial x}E_i(x)\right)w_i \quad (2.8)$$

$$\hat{f} - f = C\big(z - T(x, x_o)\big) + \sum_{i=1}^{m} DK_i(x)w_i \quad (2.9)$$

It should be noted here that the zero-input dynamics of (2.8) - (2.9) is exactly linear and moreover, if the matrix A is Hurwitz, the zero-input response is

$$z(t) - T(x(t), x_o(t)) = e^{At}\big(z(0) - T(x(0), x_o(0))\big) \to 0$$

$$\hat{f}(t) - f(t) = Ce^{At}\big(z(0) - T(x(0), x_o(0))\big) \to 0$$

which means that the manifold $z = T(x, x_o)$ is attractive and the fault estimation error $\hat{f}(t) - f(t)$ asymptotically approaches zero.

The second requirement for the residual generator is that the residual must remain completely unaffected by any disturbances $w_i(t)$ present in the system. Disturbance decoupling can be achieved if the coefficients of $w_i$ in (2.8) and (2.9) vanish, i.e, for all $i = 1,..,m$,

$$\frac{\partial T(x, x_o)}{\partial x} E_i(x) - BK_i(x) = 0 \qquad (2.10)$$

$$DK_i(x) = 0 \qquad (2.11)$$

In summary, the functional-observer-based residual generator should satisfy the following design conditions:

i) The functional observer conditions (2.6) and (2.7)

ii) The disturbance decoupling conditions (2.10) and (2.11) for all disturbances

Note that condition (2.6) is the standard invariance equation followed by all observers (whether full-state, reduced-order or functional). In effect, defining the extended vector functions

$$F_e(x, x_o) = \begin{bmatrix} F(x) + G(x)Qx_o \\ Rx_o \end{bmatrix} \qquad (2.12)$$

$$H_e(x, x_o) = H(x) + J(x)Qx_o \qquad (2.13)$$

equation (2.6) takes the classic KKL form (Andrieu & Praly, 2006; Krener & Xiao, 2018; Kazantzis & Kravaris, 1998):

$$\frac{\partial T(x, x_o)}{\partial(x, x_o)} F_e(x, x_o) = AT(x, x_o) + BH_e(x, x_o) \qquad (2.14)$$

# 3. Solution of the design conditions

For the design of the functional observer (2.4), one must be able to find matrices A, B, C and D and a differentiable map $T(x, x_o)$ so that the design conditions (2.6), (2.7), (2.10) and (2.11) are satisfied. In addition, it is desired that the matrix A is Hurwitz with fast eigenvalues for stability and fast response of the error dynamics. The following proposition provides necessary and sufficient conditions for the residual generator (2.4) to satisfy the functional observer conditions (2.6) and (2.7).

**Proposition 1:** *For the extended system* (2.3), *there exists a functional observer of the form* (2.4) *satisfying the functional observer design conditions* (2.6) *and* (2.7) *if and only if there exist constant row vectors* $v_0, v_1, ..., v_{s-1}, v_s \in \mathbb{R}^p$ *that satisfy:*

$$v_0 H_e(x, x_o) + L_{F_e}(v_1 H_e(x, x_o)) + \cdots + L_{F_e}^{s-1}(v_{s-1} H_e(x, x_o)) + L_{F_e}^s(v_s H_e(x, x_o))$$

$$+Q(R^s + \alpha_1 R^{s-1} + \cdots + \alpha_{s-1} R + \alpha_s I)x_o = 0 \qquad (3.1)$$

*where* $F_e(x, x_o)$ *and* $H_e(x, x_o)$ *as defined by (2.12) and (2,13) respectively,* $L_{F_e}$ *denotes the Lie derivative operator with respect to* $F_e$:

$$L_{F_e} = \sum_{k=1}^{n} F_k(x) \frac{\partial}{\partial x_k} + Qx_o \sum_{k=1}^{n} G_k(x) \frac{\partial}{\partial x_k} + \sum_{j=1}^{n_o} R_j x_o \frac{\partial}{\partial x_{o_j}}$$

*and* $\lambda^s + \alpha_1 \lambda^{s-1} + \cdots + \alpha_{s-1} \lambda + \alpha_s$ *is the characteristic polynomial of the matrix A.*

**Proof**: i) <u>Necessity</u>: Suppose that there exists $T(x, x_o) = \begin{bmatrix} T_1(x, x_o) \\ T_2(x, x_o) \\ \vdots \\ T_s(x, x_o) \end{bmatrix}$ such that (2.6) or (2.14) is satisfied, Using the Lie derivative notation, this condition may be written component-wise as

$$\begin{bmatrix} L_{F_e}T_1(x,x_o) \\ L_{F_e}T_2(x,x_o) \\ \vdots \\ L_{F_e}T_s(x,x_o) \end{bmatrix} = A \begin{bmatrix} T_1(x,x_o) \\ T_2(x,x_o) \\ \vdots \\ T_s(x,x_o) \end{bmatrix} + \begin{bmatrix} B_1 H_e(x,x_o) \\ B_2 H_e(x,x_o) \\ \vdots \\ B_s H_e(x,x_o) \end{bmatrix}$$

where $B_1, \ldots, B_s$ denote the rows of the matrix B. Applying the Lie derivative operator $L_{F_e}$ to each component of the above equation $(k-1)$ times, we find that for $k=1,2,3\ldots$

$$\begin{bmatrix} L_{F_e}^k T_1(x,x_o) \\ L_{F_e}^k T_2(x,x_o) \\ \vdots \\ L_{F_e}^k T_s(x,x_o) \end{bmatrix} = A^k \begin{bmatrix} T_1(x,x_o) \\ T_2(x,x_o) \\ \vdots \\ T_s(x,x_o) \end{bmatrix}$$

$$+ \begin{bmatrix} (A^{k-1}B)_1 H_e(x,x_o) + L_{F_e}((A^{k-2}B)_1 H_e(x,x_o)) + \cdots + L_{F_e}^{k-1}(B_1 H_e(x,x_o)) \\ (A^{k-1}B)_2 H_e(x,x_o) + L_{F_e}((A^{k-2}B)_2 H_e(x,x_o)) + \cdots + L_{F_e}^{k-1}(B_2 H_e(x,x_o)) \\ \vdots \\ (A^{k-1}B)_s H_e(x,x_o) + L_{F_e}((A^{k-2}B)_s H_e(x,x_o)) + \cdots + L_{F_e}^{k-1}(B_s H_e(x,x_o)) \end{bmatrix}$$

and we can calculate

$$\left(L_{F_e}^s + \alpha_1 L_{F_e}^{s-1} + \cdots + \alpha_s I\right) T_i(x,x_o) = ((A^{s-1}B)_i + \alpha_1 (A^{s-2}B)_i + \cdots + \alpha_{s-1} B_i) H_e(x,x_o)$$

$$+ L_{F_e}(((A^{s-2}B)_i + \cdots + \alpha_{s-2} B_i) H_e(x,x_o)) + \cdots + L_{F_e}^{s-1}(B_i H_e(x,x_o)) \qquad (3.2)$$

where $\alpha_1, \ldots, \alpha_s$ are the coefficients of the characteristic polynomial of the matrix A.

At the same time, $T(x,x_o)$ must satisfy (2.7), hence applying $\left(L_{F_e}^s + \alpha_1 L_{F_e}^{s-1} + \cdots + \alpha_s I\right)$ on each component of equation (2.7) and using (3.2) gives:

$(CA^{s-1}B + \alpha_1 CA^{s-2}B + \cdots + \alpha_{s-1}CB + \alpha_s D) H_e(x,x_o)$

$+ L_{F_e}((CA^{s-2}B + \cdots + \alpha_{s-2}CB + \alpha_{s-1}D) H_e(x,x_o)) + \cdots + L_{F_e}^{s-1}((CB + \alpha_1 D) H_e(x,x_o)) +$

$L_{F_e}^s(D H_e(x,x_o)) = \left(L_{F_e}^s + \alpha_1 L_{F_e}^{s-1} + \cdots + \alpha_s I\right) Q x_o = Q(R^s + \alpha_1 R^{s-1} + \cdots + \alpha_s I) x_o$

which proves that (3.1) is satisfied. □

ii) <u>Sufficiency</u>: Suppose there exist constant row vectors $v_0, v_1, \ldots, v_{s-1}, v_s$ that satisfy (3.1). Consider the following choices of (A, B, C, D) matrices:

$$A = \begin{bmatrix} 0 & 0 & \cdots & 0 & -\alpha_s \\ 1 & 0 & \cdots & 0 & -\alpha_{s-1} \\ \vdots & \ddots & \ddots & \vdots & \vdots \\ 0 & \cdots & 1 & 0 & -\alpha_2 \\ 0 & \cdots & 0 & 1 & -\alpha_1 \end{bmatrix}, \quad B = \begin{bmatrix} \alpha_s \\ \alpha_{s-1} \\ \vdots \\ \alpha_2 \\ \alpha_1 \end{bmatrix} v_s - \begin{bmatrix} v_0 \\ v_1 \\ \vdots \\ v_{s-2} \\ v_{s-1} \end{bmatrix}, \quad C = [0\ 0 \cdots 0\ 1], \quad D = -v_s \quad (3.3)$$

For the above A and C matrices (in observer canonical form), the design conditions (2.6) and (2.7) can be written component-wise as follows:

$$L_{F_e} T_1(x, x_o) + \alpha_s T_s(x, x_o) - (\alpha_s v_s - v_0) H_e(x, x_o) = 0 \tag{3.4}$$

$$L_{F_e} T_2(x, x_o) - T_1(x, x_o) + \alpha_{s-1} T_s(x, x_o) - (\alpha_{s-1} v_s - v_1) H_e(x, x_o) = 0 \tag{3.5}$$

$$\vdots$$

$$L_{F_e} T_s(x, x_o) - T_{s-1}(x, x_o) + \alpha_1 T_s(x, x_o) - (\alpha_1 v_s - v_{s-1}) H_e(x, x_o) = 0 \tag{3.6}$$

$$T_s(x, x_o) - v_s H_e(x, x_o) = Q x_o \tag{3.7}$$

We observe that the above equations are easily solvable sequentially for $T_s(x), T_{s-1}(x), \ldots, T_1(x)$, starting from the last equation and going up. In particular, for the chosen B and D matrices, we find from (3.7), (3.6), ... , (3.5):

$$T_s(x, x_o) = v_s H_e(x, x_o) + Q x_o$$

$$T_{s-1}(x, x_o) = L_{F_e}(v_s H_e(x, x_o)) + v_{s-1} H_e(x, x_o) + Q(R + \alpha_1 I) x_o$$

$$\vdots$$

$$T_2(x, x_o) = L_{F_e}^{s-2}(v_s H_e(x, x_o)) + \cdots + v_2 H_e(x, x_o) + Q(R^{s-2} + \cdots + \alpha_{s-2} I) x_o$$

$$T_1(x, x_o) = L_{F_e}^{s-1}(v_s H_e(x, x_o)) + \cdots + L_{F_e}(v_2 H_e(x, x_o)) + v_1 H_e(x, x_o)$$

$$+ Q(R^{s-1} + \cdots + \alpha_{s-1} I) x_o$$

whereas (3.4) gives:

$$L_{F_e}^s(v_s H_e(x, x_o)) + L_{F_e}^{s-1}(v_{s-1} H_e(x, x_o)) + \cdots + L_{F_e}(v_1 H_e(x, x_o)) + v_0 H_e(x, x_o)$$

$$+ Q(R^s + \cdots + \alpha_{s-1} R + \alpha_s I) x_o = 0$$

which is exactly (3.1). Thus, we have proved that

$$T(x,x_o) = \begin{bmatrix} v_1 H_e(x,x_o) + L_{F_e}(v_2 H_e(x,x_o)) + \cdots + L_{F_e}^{s-1}(v_s H_e(x,x_o)) + Q(R^{s-1} + \cdots + \alpha_{s-1}I)x_o \\ v_2 H_e(x,x_o) + \cdots + L_{F_e}^{s-2}(v_s H_e(x,x_o)) + Q(R^{s-2} + \cdots + \alpha_{s-2}I)x_o \\ \vdots \\ v_{s-1}H(x) + L_{F_e}(v_s H_e(x,x_o)) + Q(R + \alpha_1 I)x_o \\ v_s H_e(x,x_o) + Qx_o \end{bmatrix} \quad (3.8)$$

satisfies the design conditions (2.6) and (2.7) when $v_0, v_1, \ldots, v_{s-1}, v_s$ satisfy (3.1) and the A, B, C, D matrices are chosen according to (3.3). □

It is important to observe that the sufficiency part of the proof is constructive: it gives specific formulas for the (A, B, C, D) matrices of the functional observer in (3.3), as well as an explicit formula (3.8) for the solution $T(x,x_o)$ of the design equations (2.6) and (2.7), in terms of the parity vectors $v_0, v_1, \ldots, v_{s-1}, v_s$ that satisfy (3.1).

Also, it should be noted that condition (3.1) can be specialized for common fault types like bias fault (step) and drifting fault (ramp), as follows:

i) For the case of a step fault, the exo-system (2.2) is $\dot{x}_o = 0$, $f = x_o$, with $x_o$ scalar, and so

$$F_e(x,x_o) = \begin{bmatrix} F(x) + G(x)x_o \\ 0 \end{bmatrix}, H_e(x,x_o) = H(x) + J(x)x_o, \text{ and } (3.1) \text{ takes the form}$$

$$v_0(H(x) + J(x)x_o) + L_{F_e}(v_1(H(x) + J(x)x_o)) + \cdots + L_{F_e}^s(v_s(H(x) + J(x)x_o)) + \alpha_s x_o = 0$$

It should be noted that in this case, the above condition can be rescaled so as to be independent of the coefficient $\alpha_s$:

$$v_0^*(H(x) + J(x)x_o) + L_{F_e}(v_1^*(H(x) + J(x)x_o)) + \cdots + L_{F_e}^s(v_s^*(H(x) + J(x)x_o)) + x_o = 0$$

where $v_j^* = \frac{v_j}{\alpha_s}, j = 0, 1, \ldots, s$. This implies that the functional observer eigenvalues can be arbitrarily assigned to ensure stability and fast response.

ii) For the case of a ramp fault, the exo-system (2.2) is $\frac{dx_{o1}}{dt} = x_{o2}, \frac{dx_{o2}}{dt} = 0, f = x_{o1}$, i.e. it has

$$R = \begin{bmatrix} 0 & 1 \\ 0 & 0 \end{bmatrix}, Q = [1\ 0], \text{ and (3.1) takes the form}$$

$$v_0(H(x) + J(x)x_{o1}) + L_{F_e}(v_1(H(x) + J(x)x_{o1})) + \cdots + L_{F_e}^s(v_s(H(x) + J(x)x_{o1})) + \alpha_{s-1}x_{o2} + \alpha_s x_{o1} = 0$$

with the understanding that when $s = 1$, $\alpha_0 = 1$. Here functional observer eigenvalues $\lambda_j$ cannot be arbitrarily assigned; they are constrained by the ratio of $\alpha_{s-1}$ and $\alpha_s$ that satisfy the above condition: $\sum_{j=1}^{s} \frac{1}{\lambda_j} = -\frac{\alpha_{s-1}}{\alpha_s}$.

*Remark 3.1*: Because $\left(L_{F_e}^k(v_k H_e(x, x_o))\right)_{x_o=0} = L_F^k(v_k H(x))$, $k = 0,1,\cdots,s$, condition (3.1) of Proposition 1 implies that $v_0 H(x) + L_F(v_1 H(x)) + \cdots + L_F^{s-1}(v_{s-1}H(x)) + L_F^s(v_s H(x)) = 0$. This is exactly the corresponding condition in Venkateswaran et al. (2022), where only detection was sought (ibid. p. 807, eq. (3.1)). Thus, we see that the additional requirement that the residual generator's output be an estimate of the fault size makes the existence condition more restrictive.

The following Proposition provides necessary and sufficient conditions for the derived residual generator to meet the disturbance decoupling specifications (2.10) and (2.11).

**Proposition 2:** *Suppose that there exist constant row vectors $v_0, v_1, \ldots, v_{s-1}, v_s \in \mathbb{R}^p$ that satisfy (3.1) and that the matrices (A, B, C, D) have been chosen according to (3.3), so that (2.6) and (2.7) hold with $T(x, x_o)$ given by (3.8). The residual generator will satisfy the disturbance decoupling conditions (2.10) and (2.11) if and only if for all $i = 1, 2, \ldots, m$:*

$$v_{\kappa-1} K_i(x) + \sum_{\mu=\kappa}^{s} L_{E_i} L_{F_e}^{\mu-\kappa}\left(v_\mu H_e(x, x_o)\right) = 0, \qquad \kappa = 1, \cdots, s$$

$$v_s K_i(x) = 0 \qquad\qquad (3.9)$$

where $L_{E_i}$ denotes the Lie derivative operator $L_{E_i} = \sum_{j=1}^{n} E_{i_j}(x) \frac{\partial}{\partial x_j}$.

**Proof:** The disturbance decoupling conditions (2.10) and (2.11) can be written in component form, for $i = 1,2,\ldots m,$ as follows:

$$\frac{\partial T_1(x,x_o)}{\partial x} E_i(x) - B_1 K_i(x) = 0$$

$$\frac{\partial T_2(x,x_o)}{\partial x} E_i(x) - B_2 K_i(x) = 0$$

$$\vdots$$

$$\frac{\partial T_{s-1}(x,x_o)}{\partial x} E_i(x) - B_{s-1} K_i(x) = 0$$

$$\frac{\partial T_s(x,x_o)}{\partial x} E_i(x) - B_s K_i(x) = 0$$

$$D K_i(x) = 0$$

Substituting the expressions for B, D and $T(x)$ from (3.3) and (3.8) to the above equations lead to the following conditions:

$$L_{E_i} L_{F_e}^{s-1}(v_s H_e(x,x_o)) + \cdots + L_{E_i} L_{F_e}(v_2 H_e(x,x_o)) + L_{E_i}(v_1 H_e(x,x_o)) - (\alpha_s v_s - v_0) K_i(x) = 0$$

$$L_{E_i} L_{F_e}^{s-2}(v_s H_e(x,x_o)) + \cdots + L_{E_i} L_{F_e}(v_3 H_e(x,x_o)) + L_{E_i}(v_2 H_e(x,x_o)) - (\alpha_{s-1} v_s - v_1) K_i(x) = 0$$

$$\vdots$$

$$L_{E_i} L_{F_e}^2(v_s H_e(x,x_o)) + L_{E_i} L_{F_e}(v_{s-1} H_e(x,x_o)) + L_{E_i}(v_{s-2} H_e(x,x_o)) - (\alpha_3 v_s - v_{s-3}) K_i(x) = 0$$

$$L_{E_i} L_{F_e}(v_s H_e(x,x_o)) + L_{E_i}(v_{s-1} H_e(x,x_o)) - (\alpha_2 v_s - v_{s-2}) K_i(x) = 0$$

$$L_{E_i}(v_s H_e(x,x_o)) - (\alpha_1 v_s - v_{s-1}) K_i(x) = 0$$

$$-v_s K_i(x) = 0$$

which can be written equivalently as

$$L_{E_i} L_{F_e}^{s-1}(v_s H_e(x,x_o)) + \cdots + L_{E_i} L_{F_e}(v_2 H_e(x,x_o)) + L_{E_i}(v_1 H_e(x,x_o)) + v_0 K_i(x) = 0$$

$$L_{E_i}L_{F_e}^{s-2}(v_s H_e(x,x_o)) + \cdots + L_{E_i}L_{F_e}(v_3 H_e(x,x_o)) + L_{E_i}(v_2 H_e(x,x_o)) + v_1 K_i(x) = 0$$

$$\vdots$$

$$L_{E_i}L_{F_e}^2(v_s H_e(x,x_o)) + L_{E_i}L_{F_e}(v_{s-1} H_e(x,x_o)) + L_{E_i}(v_{s-2} H_e(x,x_o)) + v_{s-3} K_i(x) = 0$$

$$L_{E_i}L_{F_e}(v_s H_e(x,x_o)) + L_{E_i}(v_{s-1} H_e(x,x_o)) + v_{s-2} K_i(x) = 0$$

$$L_{E_i}(v_s H_e(x,x_o)) + v_{s-1} K_i(x) = 0$$

$$v_s K_i(x) = 0$$

This completes the proof. □

*Remark 3.2*: Because $\left(L_{F_e}^\iota(v_\mu H_e(x,x_o))\right)_{x_o=0} = L_F^\iota(v_\mu H(x))$, $\iota = 0,1,\cdots,s$, $\mu = 0,1,\cdots,s$, the condition (3.9) of Proposition 2 implies that $v_{\kappa-1}K_i(x) + \sum_{\mu=\kappa}^{s} L_{E_i}L_F^{\mu-\kappa}(v_\mu H(x)) = 0$, $\kappa = 1,\cdots,s$, leading to exactly the disturbance decoupling conditions of the residual generator in Venkateswaran et al. (2022), where only detection was sought (ibid. p. 808, eq. (3.9)).

*Remark 3.3*: In many applications, process disturbances do not affect sensors and sensor disturbances do not affect the process. Under these circumstances, the disturbance decoupling conditions (3.9) get simplified since for every disturbance, either $E_i(x)$ or $K_i(x)$ vanishes, depending on whether it is a process or sensor disturbance. A sensor disturbance generally places more restrictions than a process disturbance. In particular, we see from (3.9) that

a) A process disturbance places no restriction on $v_0$ since the corresponding $K_i(x) = 0$.

b) A sensor disturbance imposes the restriction that $[v_0\ v_1\ \ldots\ v_{s-1}\ v_s]K_i(x) = 0$. In case a disturbance affects only a specific sensor measuring $y_j$, this implies that the *j*-th element of $v_0, v_1, \ldots, v_{s-1}, v_s$ must equal to 0, hence the measurement $y_j$ must not be used in the residual generator.

The disturbance decoupled fault detection and estimation approach developed in this section can be directly extended to the case of multiple faults, so that in addition, fault isolation is accomplished. For example, in a system with $n_f$ faults, $n_f$ functional observers can be designed, one for each fault, where for functional observer $i$, fault $i$ is to be detected and all the other $n_f - 1$ faults are disturbances that are decoupled (see Figure 2).

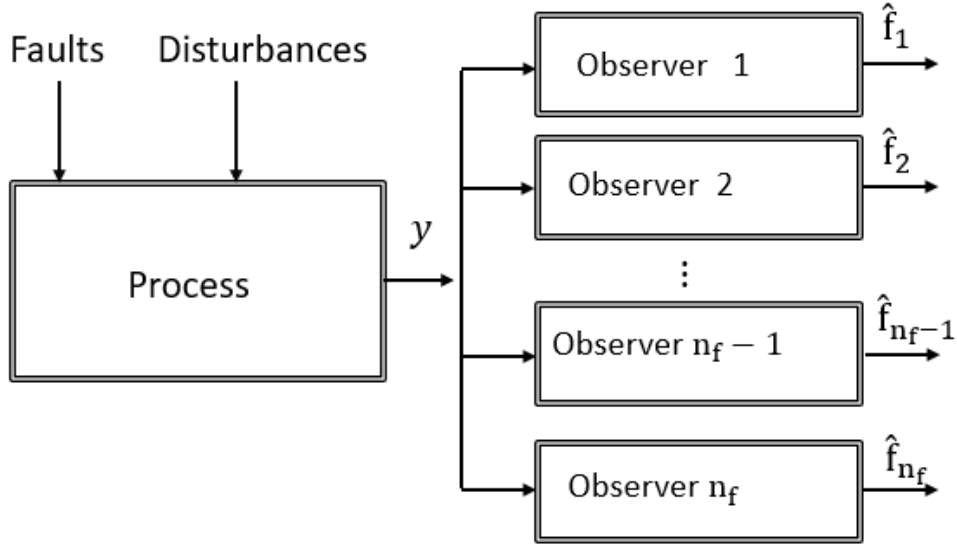

Figure 2. Fault Isolation scheme, based on a set of functional observers, one for each fault

### Detection and estimation of additive sensor faults

In the special case of only sensor faults ($G(x) = 0$) and only additive faults ($J(x) = J = constant$), condition (3.1) of Proposition 1 takes the form

$$v_0 H(x) + L_F(v_1 H(x)) + \cdots + L_F^{s-1}(v_{s-1} H(x)) + L_F^s(v_s H(x))$$
$$+ \big((v_0 J + \alpha_s)Q + (v_1 J + \alpha_{s-1})QR + \cdots + (v_{s-1} J + \alpha_1)QR^{s-1} + (v_s J + 1)QR^s\big) x_o = 0$$

which can be split into two conditions:

$$v_0 H(x) + L_F(v_1 H(x)) + \cdots + L_F^{s-1}(v_{s-1} H(x)) + L_F^s(v_s H(x)) = 0 \qquad (3.10)$$

$$(v_0 J + \alpha_s)Q + (v_1 J + \alpha_{s-1})QR + \cdots + (v_{s-1} J + \alpha_1)QR^{s-1} + (v_s J + 1)QR^s = 0 \qquad (3.11)$$

where $L_F$ denotes the Lie derivative operator $L_F = \sum_{k=1}^{n} F_k(x) \frac{\partial}{\partial x_k}$.

Condition (3.9) for disturbance decoupling (Proposition 2) takes the form

$$v_{\kappa-1}K_i(x) + \sum_{\mu=\kappa}^{s} L_{E_i} L_F^{\mu-\kappa}\left(v_\mu H(x)\right) = 0, \qquad \kappa = 1, \cdots, s$$

$$v_s K_i(x) = 0 \tag{3.12}$$

for all $i = 1, 2, ..., m$.

We see that we need conditions (3.10) and (3.12) that need to be satisfied for any residual generator, and in addition condition (3.11) in order for the residual to represent an estimate of the fault size. In the case of a step fault, this extra condition simplifies to $v_0 J + \alpha_s = 0$.

## Scalar functional observers

Sometimes it is possible to satisfy the conditions of Propositions 1 and 2 for $s = 1$, leading to a scalar residual generator. If this can happen, it would be the best situation from the point of view of ease of implementation, since the residual generator becomes a simple linear first-order filter. The conditions for $s = 1$ are as follows:

$$v_0 H_e(x, x_o) + L_{F_e}\left(v_1 H_e(x, x_o)\right) + Q(R + \alpha_1 I)x_o = 0$$

$$v_0 K_i(x) + L_{E_i}\left(v_1 H_e(x, x_o)\right) = 0, \qquad i = 1, ..., m$$

$$v_1 K_i(x) = 0, \qquad i = 1, ..., m$$

or, in terms of the original functions,

$$v_0(H(x) + J(x)Qx_o) + L_F(v_1(H(x) + J(x)Qx_o)) + [L_G(v_1(H(x) + J(x)Qx_o))]Qx_o$$

$$+ v_1 J(x)QRx_o + Q(R + \alpha_1 I)x_o = 0$$

$$v_0 K_i(x) + L_{E_i}(v_1(H(x) + J(x)Qx_o)) = 0, \qquad i = 1, ..., m$$

$$v_1 K_i(x) = 0, \qquad i = 1, ..., m$$

If the above conditions can be satisfied, the corresponding functional observer is the following:

$$\frac{dz}{dt} = -\alpha_1 z + (\alpha_1 v_1 - v_0)y$$

$$\hat{f} = z - v_1 y$$

Note that the foregoing conditions take an even simpler form in special cases:

(i) in the case of an additive sensor fault only ($G(x) = 0$, $J(x) = J = constant$), conditions become (from (3.10) – (3.12), for $s = 1$):

$$v_0 H(x) + L_F(v_1 H(x)) = 0$$

$$(v_0 J + \alpha_1)Q + (v_1 J + 1)QR = 0$$

$$v_0 K_i(x) + L_{E_i}(v_1 H(x)) = 0, \quad i = 1, \ldots, m$$

$$v_1 K_i(x) = 0, \quad i = 1, \ldots, m$$

(ii) in the case of a process fault, using sensors that are unaffected by faults or disturbances ($J(x) = 0$, $K_i(x) = 0$), conditions become:

$$v_0 H(x) + L_F(v_1 H(x)) + [L_G(v_1 H(x))]Qx_o + Q(R + \alpha_1 I)x_o \qquad (3.13)$$

$$L_{E_i}(v_1 H(x)) = 0, \quad i = 1, \ldots, m \qquad (3.14)$$

where (3.13) can be satisfied if it so happens that

$$v_0 H(x) + L_F(v_1 H(x)) = 0 \qquad (3.15)$$

$$[L_G(v_1 H(x))]Q + Q(R + \alpha_1 I) = 0 \qquad (3.16)$$

We see that in these special cases, conditions are not hard to check.

## Detection and estimation for linear systems

When the original system (2.1) is linear, i.e.

$$\dot{x} = Fx + Gf + \sum_{i=1}^{m} E_i w_i \qquad (3.17)$$

$$y = Hx + Jf + \sum_{i=1}^{m} K_i w_i$$

the design conditions (2.6), (2.7), (2.10) and (2.11) become

$$T \begin{bmatrix} F & GQ \\ 0 & R \end{bmatrix} - AT - B[H \quad JQ] = 0 \tag{3.18}$$

$$CT + D[H \quad JQ] = [0 \quad Q] \tag{3.19}$$

$$T \begin{bmatrix} E \\ 0 \end{bmatrix} - BK = 0 \tag{3.20}$$

$$DK = 0 \tag{3.21}$$

where $E = [E_1 \ ... \ E_m]$ and $K = [K_1 \ ... \ K_m]$.

For the choices of A,B,C,D matrices given by (3.3),

$$T = \begin{bmatrix} v_1[H \quad JQ] + v_2[H \quad JQ]\begin{bmatrix} F & GQ \\ 0 & R \end{bmatrix} + \cdots + v_s[H \quad JQ]\begin{bmatrix} F & GQ \\ 0 & R \end{bmatrix}^{s-1} + [0 \quad Q(R^{s-1} + \cdots + \alpha_{s-1}I)] \\ v_2[H \quad JQ] + \cdots + v_s[H \quad JQ]\begin{bmatrix} F & GQ \\ 0 & R \end{bmatrix}^{s-2} + [0 \quad Q(R^{s-2} + \cdots + \alpha_{s-2}I)] \\ \vdots \\ v_{s-1}[H \quad JQ] + v_s[H \quad JQ]\begin{bmatrix} F & GQ \\ 0 & R \end{bmatrix} + [0 \quad Q(R + \alpha_1 I)] \\ v_s[H \quad JQ] + [0 \quad Q] \end{bmatrix}$$

and the conditions on the residual generator can be combined in a compact form as

$$[v_0 \quad v_1 \quad ... \quad v_{s-1} \quad v_s][\Gamma_o \quad \Gamma_f \quad \Gamma_w] = [0 \quad -QP_A(R) \quad 0] \tag{3.22}$$

where

$$\Gamma_o = \begin{bmatrix} H \\ HF \\ \vdots \\ HF^{s-1} \\ HF^s \end{bmatrix}, \Gamma_f = \begin{bmatrix} J & 0 & \cdots & 0 & 0 \\ HG & J & \cdots & 0 & 0 \\ HFG & HG & \ddots & \vdots & \vdots \\ \vdots & \vdots & \cdots & J & 0 \\ HF^{s-1}G & HF^{s-2}G & \cdots & HG & J \end{bmatrix} \begin{bmatrix} Q \\ QR \\ \vdots \\ QR^{s-1} \\ QR^s \end{bmatrix}, \Gamma_w = \begin{bmatrix} K & 0 & \cdots & 0 & 0 \\ HE & K & \cdots & 0 & 0 \\ \vdots & \vdots & \ddots & \vdots & \vdots \\ HF^{s-2}E & HF^{s-3}E & \cdots & K & 0 \\ HF^{s-1}E & HF^{s-2}E & \cdots & HE & K \end{bmatrix}$$

and $P_A(R) = R^s + \cdots + \alpha_{s-1}R + \alpha_s I$.

*Remark 3.4*: Condition (3.22) is very similar to the one derived in Venkateswaran et al. (2022) for detection only (ibid. p. 811, eq. (3.17)), but now $\Gamma_f$ depends on the characteristics of the exo-system.

## Functional observer with nonlinear output injection

The linear functional observer formulation developed in the previous and the present section is amenable to a slight generalization. Instead of using a linear functional observer (2.4) as residual generator, it is possible to use one involving additive nonlinear output injection terms:

$$\dot{z} = Az + \beta(y) \qquad (3.23)$$

$$\hat{f} = Cz + \delta(y)$$

where $\beta(\cdot)$ and $\delta(\cdot)$ are smooth functions $\mathbb{R}^p \to \mathbb{R}^s$ and $\mathbb{R}^p \to \mathbb{R}$ respectively. Then, the mapping $T(x, x_o)$ from $\mathbb{R}^{n+n_o}$ to $\mathbb{R}^s$ must satisfy the functional observer conditions:

$$\frac{\partial T(x, x_o)}{\partial x}(F(x) + G(x)Qx_o) + \frac{\partial T(x, x_o)}{\partial x_o}Rx_o = AT(x, x_o) + \beta(H(x) + J(x)Qx_o) \qquad (3.24)$$

$$CT(x, x_o) + \delta\big((H(x) + J(x)Qx_o)\big) = Qx_o \qquad (3.25)$$

and the observer error dynamics, in the absence of faults and disturbances, will still be linear:

$$\frac{d(z - T(x, x_o))}{dt} = A(z - T(x, x_o))$$

$$\hat{f} - f = C(z - T(x, x_o))$$

therefore it will have the same convergence properties when A is Hurwitz. Proposition 1 is generalized as follows:

**Proposition 1':** *There exists a residual generator of the form* (3.23) *satisfying the functional observer design conditions* (3.24) *and* (3.25) *if and only if there exist functions* $v_0(y)$, $v_1(y)$, ..., $v_{s-1}(y)$, $v_s(y)$ *from $\mathbb{R}^p$ to $\mathbb{R}$ that satisfy:*

$$v_0\big(H_e(x, x_o)\big) + L_{F_e}\big(v_1(H_e(x, x_o))\big) + \cdots + L_{F_e}^{s-1}\big(v_{s-1}(H_e(x, x_o))\big) + L_{F_e}^s\big(v_s(H_e(x, x_o))\big)$$

$$+ Q(R^s + \alpha_1 R^{s-1} + \cdots + \alpha_{s-1} R + \alpha_s I)x_o = 0 \qquad (3.26)$$

If such functions can be found, the functional observer can be built by using

$$A = \begin{bmatrix} 0 & 0 & \cdots & 0 & -\alpha_s \\ 1 & 0 & \cdots & 0 & -\alpha_{s-1} \\ \vdots & \ddots & \ddots & \vdots & \vdots \\ 0 & \cdots & 1 & 0 & -\alpha_2 \\ 0 & \cdots & 0 & 1 & -\alpha_1 \end{bmatrix}, \beta(y) = \begin{bmatrix} \alpha_s \\ \alpha_{s-1} \\ \vdots \\ \alpha_2 \\ \alpha_1 \end{bmatrix} v_s(y) - \begin{bmatrix} v_0(y) \\ v_1(y) \\ \vdots \\ v_{s-2}(y) \\ v_{s-1}(y) \end{bmatrix}, C = [0\ 0 \cdots 0\ 1], \delta(y) = -v_s(y) \quad (3.27)$$

and it is possible to verify that

$$T(x, x_o) = \begin{bmatrix} v_1(H_e(x, x_o)) + L_{F_e}(v_2(H_e(x, x_o))) + \cdots + L_{F_e}^{s-1}(v_s(H_e(x, x_o))) + Q(R^{s-1} + \cdots + \alpha_{s-1}I)x_o \\ v_2(H_e(x, x_o)) + \cdots + L_{F_e}^{s-2}(v_s(H_e(x, x_o))) + Q(R^{s-2} + \cdots + \alpha_{s-2}I)x_o \\ \vdots \\ v_{s-1}(H_e(x, x_o)) + L_{F_e}(v_s(H_e(x, x_o))) + Q(R + \alpha_1 I)x_o \\ v_s(H_e(x, x_o)) + Qx_o \end{bmatrix}$$

satisfies (3.24) and (3.25).

One can accordingly derive disturbance decoupling conditions for the residual generator, which will impose extra conditions on the functions $v_0(y)$, $v_1(y)$, ..., $v_{s-1}(y)$, $v_s(y)$, mutatis mutandis.

*Remark 3.5*: Because the functional observer (3.23) is a generalization of (2.4), the conditions are less restrictive, but they are harder to check. However, in special cases, they might be useful. Suppose for example, that we wish to build a scalar residual generator for a process fault, using sensors unaffected by faults or disturbances, but conditions (3.13) and/or ((3.14) cannot hold for constant $v_0, v_1$. Then, we might try a scalar residual generator of the form (3.23), for which the conditions for the functions $v_0(y)$, $v_1(y)$ are:

$$v_0(H(x)) + L_F(v_1(H(x))) + [L_G(v_1(H(x)))]Qx_o + Q(R + \alpha_1 I)x_o \quad (3.28)$$

$$L_{E_i}(v_1(H(x))) = 0, \quad i = 1, \ldots, m \quad (3.29)$$

where (3.28) can be satisfied if it so happens that

$$v_0(H(x)) + L_F(v_1(H(x))) = 0 \quad (3.30)$$

$$[L_G(v_1(H(x)))]Q + Q(R + \alpha_1 I) = 0 \quad (3.31)$$

# 4. Application: Fault detection and estimation for a safety-critical chemical reactor

Liquid-phase oxidation reactions are notorious for being highly exothermic and for involving serious safety threats. One well-studied example is the reaction of N-methyl pyridine (A) with hydrogen peroxide (B) in the presence of a catalyst. The product of the reaction, Methyl Pyridine N-Oxide is an important intermediate in several reactions in pharmaceutical industry (Cui et al., 2015). The key issue in the operation of liquid-phase oxidation reactors is safety: both the organic reactant is usually hazardous at high temperatures, and also hydrogen peroxide decomposition could pose serious safety threats.

It is assumed the reactor is well-mixed and has constant volume with an inlet stream containing N-methyl pyridine (A) + catalyst Z (assumed to be fully dissolved) and hydrogen peroxide (B). The catalyst is assumed to be completely dissolved in the pyridine stream and its concentration is assumed to be constant in the reactor The dynamics of the reactor (Cui et al., 2015), including possible faults and disturbances, is described by:

$$\frac{dc_A}{dt} = \frac{F}{V}(c_{A_{in}} - c_A) - \mathcal{R}(c_A, c_B, \theta, w) \tag{4.1}$$

$$\frac{dc_B}{dt} = \frac{F}{V}(c_{B_{in}} - c_B) - \mathcal{R}(c_A, c_B, \theta, w)$$

$$\frac{d\theta}{dt} = \frac{F}{V}(\theta_{in} - \theta) + \frac{(-\Delta H)_R}{\rho c_p}\mathcal{R}(c_A, c_B, \theta, w) - \frac{UA}{\rho c_p V}(\theta - \theta_J)$$

$$\frac{d\theta_J}{dt} = \frac{F_J}{V_J}(\theta_{J_{in}} + f_2 - \theta_J) + \frac{UA}{\rho_J c_{p_J} V_J}(\theta - \theta_J)$$

with measured outputs

$$y_1 = c_A + f_1$$

$$y_2 = \theta$$

$$y_3 = \theta_J$$

In the above model, $c_A$ and $c_B$ are the concentrations of Pyridine and Hydrogen Peroxide respectively in the reacting mixture, $\theta$ and $\theta_J$ are the temperatures of the reacting mixture and the jacket fluid respectively; these are the system states. $f_2$ is a possible step-like fault originating from a potential malfunction of the coolant feeding system and $f_1$ represents a possible drifting fault (ramp-type) in the analytical sensor. The reaction rate $\mathcal{R}(c_A, c_B, \theta, w)$ is an empirically derived expression and is given by $\mathcal{R}(c_A, c_B, \theta, w) = \dfrac{A_1 e^{-\frac{E_1}{\theta}} A_2 e^{-\frac{E_2}{\theta}} c_A c_B Z}{1 + A_2 e^{-\frac{E_2}{\theta}} c_B} + A_3 e^{-\frac{E_3}{\theta}} c_A c_B + w$, in which $w$ represents the uncertainty in the empirical fitting of the reaction rate expression. $c_{A_{in}}$ and $c_{B_{in}}$ are the feed concentrations of species Pyridine and Hydrogen Peroxide respectively, F and $F_J$ are the feed and coolant flowrates respectively, $\theta_{in}$ and $\theta_{J,in}$ are the inlet temperatures for the reactor and the cooling jacket respectively, V and $V_J$ are the reactor volume and cooling jacket volume respectively, $(-\Delta H)_R$ is the heat of reaction, $\rho, c_p$ and $\rho_J, c_{p_J}$ are the densities and heat capacities of the reactor contents and cooling fluid respectively, U and A are the overall heat transfer coefficient and heat transfer area respectively. $A_1, A_2, A_3$ and $E_1, E_2, E_3$ are the reaction rate parameters, pre-exponential factors and rescaled activation energies respectively, and Z is the catalyst concentration (constant).

When reactants are potentially hazardous, special precautions are taken in terms of using relatively dilute feeds and the reaction taking place at a relatively low temperature.

The goal is to design a fault diagnosis scheme that can detect, isolate and estimate these two possible faults $f_1$ and $f_2$ in the presence of uncertainty in the reaction rate. To this end, two scalar functional observers are built (i) to estimate the analytical sensor fault ($f_1$) while considering $f_2$ as an additional disturbance. (ii) to estimate inlet jacket temperature fault ($f_2$) considering $f_1$ as an additional disturbance.

To derive the functional observers, it is convenient to perform appropriate translation of axes to shift the equilibrium point to the origin by defining

$$c'_A = c_A - c_{A,s}, c'_B = c_B - c_{B,s}, \theta' = \theta - \theta_s, \theta'_J = \theta_J - \theta_{J,s},$$

where $(c_{A,s}, c_{B,s}, \theta_s, \theta_{J,s})$ is the steady state (equilibrium point) of the reactor. Then the dynamic model takes the form:

$$\frac{dc'_A}{dt} = -\frac{F}{V}c'_A - [\Re(c'_A + c_{A,s}, c'_B + c_{B,s}, \theta' + \theta_s, w) - \Re(c_{A,s}, c_{B,s}, \theta_s)] \quad (4.2)$$

$$\frac{dc'_B}{dt} = -\frac{F}{V}c'_B - [\Re(c'_A + c_{A,s}, c'_B + c_{B,s}, \theta' + \theta_s, w) - \Re(c_{A,s}, c_{B,s}, \theta_s)]$$

$$\frac{d\theta'}{dt} = -\frac{F}{V}\theta' + \frac{(-\Delta H)_R}{\rho c_p}[\Re(c'_A + c_{A,s}, c'_B + c_{B,s}, \theta' + \theta_s, w) - \Re(c_{A,s}, c_{B,s}, \theta_s)] - \frac{UA}{\rho c_p V}(\theta' - \theta'_J)$$

$$\frac{d\theta'_J}{dt} = \frac{F_J}{V_J}(f_2 - \theta'_J) + \frac{UA}{\rho_J c_{p_J} V_J}(\theta' - \theta'_J)$$

$$y'_1 = c'_A + f_1$$

$$y'_2 = \theta'$$

$$y'_3 = \theta'_J$$

In what follows: (i) Functional Observer 1 is built to estimate the analytical sensor fault ($f_1$) considering the other fault $f_2$ and the reaction rate uncertainty $w$ as disturbances; (ii) Functional Observer 2 is built to estimate inlet jacket temperature fault ($f_2$) considering the other fault $f_1$ and the reaction rate uncertainty $w$ as disturbances.

<u>Functional Observer 1</u>: The analytical sensor fault ($f_1$) is considered to be a drifting fault (ramp), hence it is generated by the exo-system defined by the equations $\frac{dx_{o1}}{dt} = x_{o2}, \frac{dx_{o2}}{dt} = 0, f_1 = x_{o1}$, and these are appended to the process system (4.2)

A scalar ($s = 1$) functional observer is sought. Checking the pertinent functional observer and disturbance decoupling conditions from Propositions 1 and 2, we see that they can only be satisfied for $\alpha_1 = \frac{F}{V}$, with corresponding parity vectors $\upsilon_0$ and $\upsilon_1$ given by:

$$\upsilon_0 = \left[ -\frac{F}{V} \quad -\frac{\rho c_p}{(-\Delta H_R)}\left(\frac{F}{V} + \frac{UA}{\rho c_p V}\right) \quad \frac{UA}{(-\Delta H_R)V} \right] \quad (4.3)$$

$$\upsilon_1 = \left[ -1 \quad -\frac{\rho c_p}{(-\Delta H_R)} \quad 0 \right]$$

The scalar functional observer is built according to (3.3), with the following design parameters:

$$A = -\alpha_1 = -\frac{F}{V}, \quad B = \alpha_1 \upsilon_1 - \upsilon_0, \quad C = 1, \quad D = -\upsilon_1$$

therefore it is given by

$$\frac{dz}{dt} = -\frac{F}{V}z + \frac{UA}{(-\Delta H_R)V}y'_2 - \frac{UA}{(-\Delta H_R)V}y'_3$$

$$\hat{f}_1 = z + y'_1 + \frac{\rho c_p}{(-\Delta H_R)}y'_2 \quad (4.4)$$

<u>Functional Observer 2</u>: The inlet cooling jacket temperature fault ($f_2$) is considered to be a bias (step-like) fault, hence it is generated by the exo-system defined by the equation $\frac{df_2}{dt} = 0$ and this is appended to system (4.2)

A scalar ($s = 1$) functional observer can be designed with eigenvalue assignment, satisfying all conditions of Propositions 1 and 2, with the following choices of parity vectors $\upsilon_0$ and $\upsilon_1$:

$$\upsilon_0 = -\alpha_1 \left[0, -\frac{UA}{\rho_j C_{pj} F_j}, 1 + \frac{UA}{\rho_j C_{pj} F_j}\right] \tag{4.5}$$

$$\upsilon_1 = -\alpha_1 \left[0, 0, \frac{V_j}{F_j}\right]$$

The scalar functional observer is built according to (3.3), with design parameters

$$A = -\alpha_1, \quad B = \alpha_1 \upsilon_1 - \upsilon_0, \quad C = 1, \quad D = -\upsilon_1$$

therefore it is given by

$$\frac{dz}{dt} = -\alpha_1 z - \left(\alpha_1 \frac{UA}{\rho_j c_{p_j} F_j}\right) y_2' - \alpha_1 \left(\alpha_1 \frac{V_j}{F_j} - 1 - \frac{UA}{\rho_j C_{pj} F_j}\right) y_3'$$

$$\hat{f}_2 = z + \alpha_1 \frac{V_j}{F_j} y_3' \tag{4.6}$$

The value of $\alpha_1 = 0.01$ was used in the simulations.

For the following process parameter values (see Cui et al., 2015):

$c_{A,in} = 4 \frac{mol}{l}$, $c_{B,in} = 3 \frac{mol}{l}$, $\theta_{in} = 333$ K, $\theta_{J,in} = 300$ K, $F = 0.02 \frac{1}{min}$, $F_j = 1 \frac{1}{min}$,

$V = 1\,\text{l},\ V_J = 3 \times 10^{-2}\,\text{l},\ A_1 = e^{8.08}\,\text{l mol}^{-1}\text{s}^{-1},\ A_2 = e^{28.12}\,\text{l mol}^{-1}\text{s}^{-1},\ A_3 = e^{25.12}\,\text{l mol}^{-1},$

$E_1 = 3952\,\text{K},\ E_2 = 7927\,\text{K},\ E_3 = 12989\,\text{K},\ \Delta H_R = -160\,\frac{\text{kJ}}{\text{mol}},$

$\rho = 1200\,\frac{\text{g}}{\text{l}},\ \rho_J = 1200\,\frac{\text{g}}{\text{l}},\ c_{p_J} = 3.4\,\frac{\text{J}}{\text{gK}},\ c_p = 3.4\,\frac{\text{J}}{\text{gK}},\ UA = 0.942\,\frac{\text{W}}{\text{K}},\ Z = 0.0021\,\frac{\text{mol}}{\text{l}}$

the corresponding reactor steady state, in the absence of faults or disturbances, is:

$c_{A,s} = 1.211\,\frac{\text{mol}}{\text{l}},\ c_{B,s} = 0.211\,\frac{\text{mol}}{\text{l}},\ \theta_s = 386.20\,\text{K},\ \theta_{J,s} = 300.02\,\text{K},$

The two functional observers are simulated for the following scenario:

Two persistent faults $f_1(t) = \begin{cases} 0, & t < 2000\text{s} \\ 0.001t - 1, & t \geq 2000\text{s} \end{cases}$, $f_2(t) = \begin{cases} 0, & t < 5000\text{s} \\ 10, & t \geq 5000\text{s} \end{cases}$ are assumed to occur along with the disturbance $w(t) = 10^5$. It is assumed that initially the process is operating at steady state. An initialization error $z(0) - T(x(0), x_o(0)) = 1$ is assumed for both observers (where $T(x, x_o)$ is given by equation (3.8)). The fault estimates are plotted in Figure 3.

Both estimates from time $t = 0$ to 2000s are decay to 0 in the absence of faults. When the sensor fault occurs at time $t=2000$s a deviation is seen in $\hat{f}_1$ whereas $\hat{f}_2$ is equal to 0. At time $t=5000$s a deviation is observed in $\hat{f}_2$ indicating the presence of a fault in thse inlet coolant temperature. Both profiles eventually converge to their actual fault profiles.

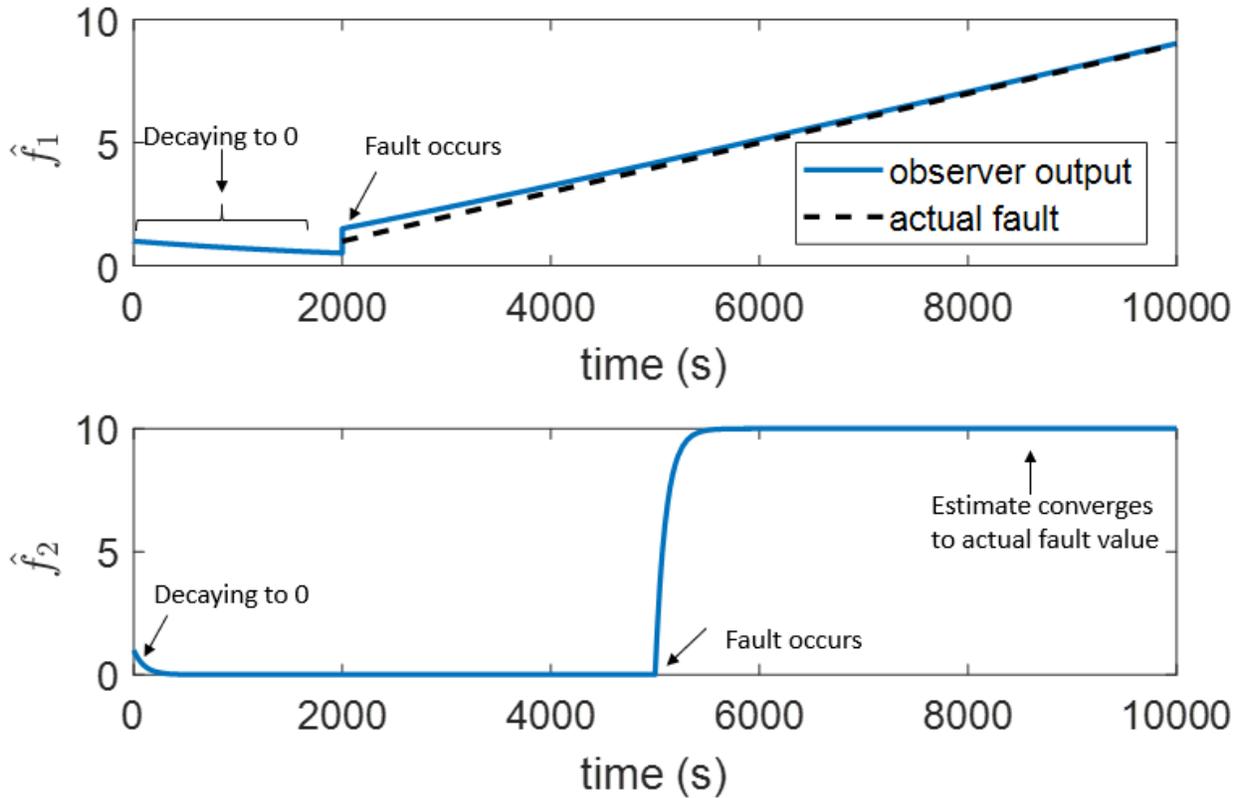

Figure 3: Fault estimates vs time with initialization error=1. The fault estimates converge to their actual trajectories

## 5. Conclusions

The problem of designing linear residual generators, where the residual signal represents an estimate of the fault, has been studied. Necessary and sufficient conditions for the pertinent functional observer, including disturbance decoupling conditions, have been derived. The conditions are more restrictive than the ones in Venkateswaran et al. (2022) where only detection was sought, because of the additional requirement of the present formulation that the residual generator's output is an estimate of the fault size. The results have been successfully applied to a chemical reactor case study.

## 6. Acknowledgment

Financial support from the National Science Foundation through the grant CBET-2133810 is gratefully acknowledged.